\documentclass{WileyMSP-template}

\usepackage{pdfpages}
\usepackage{pgffor}

\usepackage{xcolor}
\usepackage[sorting=none]{biblatex}
\begin{document}

\pagestyle{fancy}
\rhead{\includegraphics[width=2.5cm]{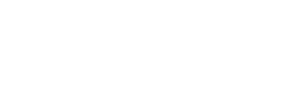}}

\title{Unexpected Anisotropic Mn-Sb Anti-site Distribution and Van der Waals Epitaxy of MnSb\textsubscript{2}Te\textsubscript{4}}

\maketitle


\author{Gustavo Chavez Ponce de Leon*,}
\author{Ahmad Dibajeh,}
\author{Gert ten Brink,}
\author{Majid Ahmadi,}
\author{Bart Jan Kooi,}
\author{George Palasantzas*}

\begin{affiliations}
MSc. Gustavo Chavez Ponce de Leon\\
Address: Zernike Institute for Advanced Materials, University of Groningen,  9747 AG Groningen, The Netherlands\\
Email Address: g.ponce.de.leon@rug.nl\\
 \vspace{15pt} 
MSc. Ahmad Dibajeh\\
Address: Zernike Institute for Advanced Materials, University of Groningen,  9747 AG Groningen, The Netherlands\\
Email Address: a.j.f.dibajeh@rug.nl\\
 \vspace{15pt} 
Dr. Gert ten Brink\\
Address: Zernike Institute for Advanced Materials, University of Groningen,  9747 AG Groningen, The Netherlands\\
Email Address: g.h.ten.brink@rug.nl\\
 \vspace{15pt} 
Dr. Majid Ahmadi\\
Addresses: (1) Zernike Institute for Advanced Materials, University of Groningen,  9747 AG Groningen, The Netherlands. (2) CogniGron (Groningen Cognitive Systems and Materials Center), University of Groningen, 9747 AG Groningen, Netherlands.\\
Email Address: majid.ahmadi@rug.nl\\
 \vspace{15pt} 
Prof. Dr. Ir. Bart Jan Kooi\\
Addresses: (1) Zernike Institute for Advanced Materials, University of Groningen,  9747 AG Groningen, The Netherlands. (2) CogniGron (Groningen Cognitive Systems and Materials Center), University of Groningen, 9747 AG Groningen, Netherlands.\\
Email Address: b.j.kooi@rug.nl\\
 \vspace{15pt} 
Prof. Dr. George Palasantzas\\
Addresses: (1) Zernike Institute for Advanced Materials, University of Groningen,  9747 AG Groningen, The Netherlands. (2) CogniGron (Groningen Cognitive Systems and Materials Center), University of Groningen, 9747 AG Groningen, Netherlands.\\
Email Address: g.palasantzas@rug.nl\\
\end{affiliations}


\keywords{Pulsed Laser Deposition [PLD], MnSb2Te4, Magnetic Topological Insulator, Van der Waals Epitaxy, Scanning Transmission Electron Microscopy [STEM], Anti-site Defects, Anisotropy}

\begin{abstract}
\justifying
Mn-Sb site mixing directly impacts both the magnetic and topological properties of MnSb\textsubscript{2}Te\textsubscript{4}. This study reveals, unlike previously believed, that these anti-sites can be unevenly distributed within the crystal. To that end, a polycrystalline sample was created with a two-step synthesis using MnTe and Sb\textsubscript{2}Te\textsubscript{3} as precursors. DC-SQUID magnetometry was used to confirm its magnetic properties. In addition, the use of High-Resolution Scanning Transmission Electron Microscopy combined with Energy-Dispersive X-ray Spectroscopy allowed us to identify the presence of an inversion-breaking asymmetry in the anti-site distribution. This reduced-symmetry structure bears resemblance to the recently proposed class of Janus materials and thus warrants further exploration due to its potential for combining topology and magnetism with other effects, such as non-linear optics and piezoelectricity. Finally, to further elucidate the interplay between site mixing, doping, topology, and magnetism, a method for growing MnSb\textsubscript{2}Te\textsubscript{4} thin films over amorphous SiO\textsubscript{x} using Sb\textsubscript{2}Te\textsubscript{3} seeds is introduced. The successful Van der Waals epitaxy of MnSb\textsubscript{2}Te\textsubscript{4} over Sb\textsubscript{2}Te\textsubscript{3} seeds using Pulsed Laser Deposition is confirmed using Scanning Transmission Electron Microscopy. This represents a crucial step in incorporating these materials into a Si-based architecture, which offers the possibility of controlling the Fermi lever via gating.
\end{abstract}


\newpage

\section{Introduction}
\justifying

In the last decades, chalcogenides-based topological insulators [TIs] (e.g. Bi\textsubscript{2}Se\textsubscript{3}, Bi\textsubscript{2}Te\textsubscript{3}, Sb\textsubscript{2}Te\textsubscript{3}, and their alloys) have become a promising class of materials for applications in spintronics, optoelectronics, and quantum computing \cite{heremans2017tetradymites,tokura2019magnetic,ortmann2015topological,he2022topological}. While their defining feature of an insulating bulk with conductive surface states is of great interest on its own right \cite{ortmann2015topological,hasan2010colloquium,qi2008topological}; the properties of TIs are significantly bolstered when their surfaces are interfaced with materials exhibiting other physical qualities. For instance, the proximity to superconductors \cite{hasan2010colloquium,he2015weak}, ferromagnets \cite{tokura2019magnetic,he2015quantum,richardella2015topological} or even other insulators \cite{murakami2015hybridization}, has been linked to exotic phenomena like the the elusive Majorana modes \cite{fu2009probing, kayyalha2020absence}, the Quantum Anomalous Hall Effect [QAHE] \cite{yu2010quantized, chang2013experimental}, and Weyl states \cite{burkov2011weyl,armitage2018weyl}.\vspace{15pt} 

The close proximity of TIs with these materials, however, is not a trivial enterprise as chalcogenides show an incredibly rich surface chemistry \cite{spataru2014fermi,walsh2017interface,shaughnessy2014energetics}. This, unfortunately, works to the detriment of their qualities, thus hindering the maximum benefits of TIs in current applications. For instance, common electrode like Au or Pt have been reported to react with many TIs creating uncontrolled intermixing layers \cite{walsh2017interface,shaughnessy2014energetics}, while the interface with Fe, Co or Ni ---standard ferromagnets--- has been shown to create a "dead" intermixing layer with neither topological nor magnetic properties \cite{walsh2017interface,gupta2009interface,majumder2017interfacial,li2012magnetic}. \vspace{15pt}

This has shifted the paradigm to search for TIs that naturally combine the presence of topological states with other physical features \cite{tokura2019magnetic}. For instance, intrinsically magnetic TIs have successfully been created by modulated doping of (Bi,Sb)\textsubscript{2}Te\textsubscript{3} with Cr and V; thereby enabling the observation of the AHE and axion states \cite{tokura2019magnetic,mogi2017tailoring,chen2010massive}. Nevertheless, this approach is soley suitable for devices operating at exceedingly low temperatures. Other materials, which inherently incorporate the magnetic atoms into a stoichiometric crystal structure (e.g. MnBi\textsubscript{2}Te\textsubscript{4} \cite{otrokov2019prediction}, MnSb\textsubscript{2}Te\textsubscript{4} \cite{yan2019evolution,murakami2019realization,ge2021direct, liu2021site,wimmer2021mn,folkers2022occupancy,xi2022relationship}, Mn(Bi,Sb)\textsubscript{2}Te\textsubscript{4} \cite{chen2020ferromagnetism}, etc.), have successfully preserved the magnetic order to higher temperatures without sacrificing the signatures of topology. \vspace{15pt} 

\begin{figure}
  \includegraphics[width=\linewidth]{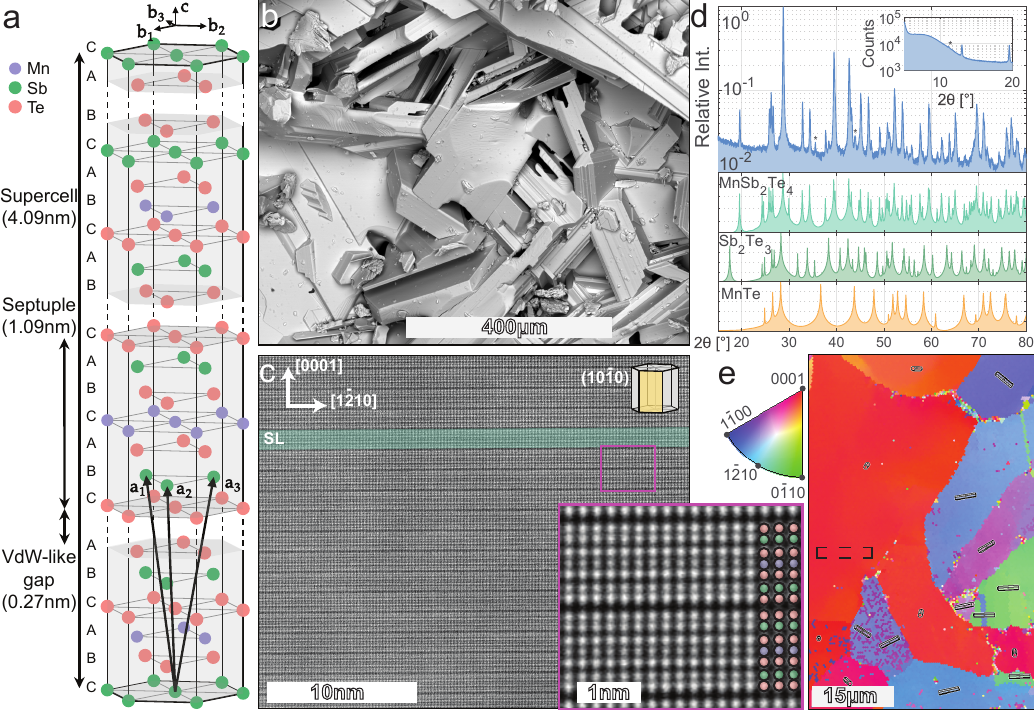}
  \caption{Characterization of polycrystalline MnSb\textsubscript{2}Te\textsubscript{4} sample. a) Schematic representation of the ideal crystal structure. The a-vectors indicate the primitive rhombohedral cell, while the b-vectors represent the hexagonal supercell. b) CBS SEM micrograph of the ingot as-synthesized, the crystallites have a considerable size of a couple hundreds of microns exhibiting sharp angular features. No compositional contrast is observed indicative of the ingot's uniformity. c) HAADF STEM images of a single grain. The crystal consists of perfectly arranged SL separated by a Van der Waals-like gap. The inset shows a zoom in version of the SL captured using drift corrected frame integration [DCFI] and post-processed with a radial Wiener filter. The difference in contrast due to atomic number clearly highlights the location of Mn in the structure. d) XRD of the polycrystalline sample compared with the simulated pattern of MnSb\textsubscript{2}Te\textsubscript{4} (as reported by \cite{orujlu2021phase}) and its precursors (Sb\textsubscript{2}Te\textsubscript{3} \cite{anderson1974refinement} and MnTe \cite{kruger1967lattice}). The asterisk indicate background peaks. e) Inverse pole figure of the gridded ingot obtained using EBSD. The selected area indicates the approximate location where the lamella was taken.}
  \label{fig:Ingot}
\end{figure}

From this group, MnSb\textsubscript{2}Te\textsubscript{4} [MST] stands out as one of the TIs with a ferromagnetic order with some of the highest reported Curie temperature \cite{wimmer2021mn,folkers2022occupancy}. Ideal MST possesses a trigonal $R\bar{3}m$ (tetradymite) structure, which can be described using a hexagonal unit cell with an A,B,C stacking sequence of close packed planes where covalently boned septuple layers [SL: Te-Sb-Te-Mn-Te-Sb-Te] are separated by a van der Waals [vdW]-like gap \cite{eremeev2017competing,kooi2020chalcogenides} (see  \textbf{Figure \ref{fig:Ingot}a}). Similar to its Bi analog (MnBi\textsubscript{2}Te\textsubscript{4}), ideal MST is predicted to be a TI with an antiferromagnetic order {composed of ferromagnetically ordered Mn planes with an antiferromagnetic interplane coupling where all the magnetic moments align parallel to the crystallographic c-axis} \cite{eremeev2017competing,wimmer2021mn}. In practice, however, not only has the antiferromagnetic order been observed, but also a ferromagnetic interplane one \cite{murakami2019realization,chen2020ferromagnetism,liu2021site,ge2021direct,wimmer2021mn}. This property was quickly linked to the presence of Mn-Sb site mixing \cite{murakami2019realization,liu2021site}, which shows significant variation depending on synthesis conditions, thereby explaining the diversity in reported properties. \vspace{15pt} 

While the observation of Mn-Sb anti-sites represents an important advancement in understanding MST, not all of its magnetic properties have been fully explained. Indeed, magnetization measurements in the low-temperature region are often accompanied by 'bumps' and 'kinks' irrespective of the synthesis method \cite{chen2020ferromagnetism,yan2019evolution,liu2021site,wimmer2021mn,xi2022relationship,folkers2022occupancy} strongly suggesting deviations from ideal ferromagnetism. Moreover, even in the paramagnetic phase, MST exhibits deviations from the Curie-Weiss law not present in other similar alloys (e.g., MnBi\textsubscript{2-x}Sb\textsubscript{x}Te\textsubscript{4})\cite{yan2019evolution,murakami2019realization}. Here, the high-temperature behavior ---consistent with an antiferromagnetic order --- deviates below $\sim$50K-80K to follow a Curie-Weiss law reminiscent of a ferromagnet.\vspace{15pt}

In addition to modifying the magnetic properties, these anti-sites also influence the topological and electrical properties of MST. Indeed, Mn acts as a p-type electrical dopant\cite{qin2019achieving, levy2023high} which, in conjunction with the well-reported tendency of Sb and Te to create heavy p-doped crystals \cite{hsieh2009observation,wang2010atomically,heremans2017tetradymites,suh2015probing}, further lowers the Fermi-level away from the bulk bandgap and into the valence band, thus hindering the observation of the exotic properties associated with band topology.\vspace{15pt}

Given that Mn-Sb site mixing governs many of the relevant properties of MST, the need to study the specific spatial distribution of these anti-sites was quickly pointed out \cite{chen2020ferromagnetism}. The study by Y. Liu et al. addressed the question using neutron diffraction, where the absence of structured diffuse scattering patterns was indicative of a random distribution of defects, albeit limited by the probe resolution of $\sim$1nm \cite{liu2021site}. \vspace{15pt}

In this work, we show that the anti-sites can actually develop an inversion-symmetry-breaking distribution. Our recent Scanning Transmission Electron Microscopy [STEM] results indicate that the Mn-Sb site mixing has a preferential direction, distinguishing between the two Sb planes, where one of the Sb layers presents a higher Mn concentration. The asymmetry is only observed  within single septuple layers ($<$1nm), therefore explaining its absence in neutron scattering experiments. Moreover, the use of Superconductive Quantum Interference Device [SQUID] magnetometry revealed that our sample exhibits hysteresis and remanent magnetization consistent with the ferrimagnetic order.\vspace{15pt}

Finally, in order to address the previously mentioned problem of intrinsic p-doping in as-grown MST, we present a method for growing MST films over amorphous substrates, such as SiO\textsubscript{x}, which enhances silicon wafer compatibility and thus offers the possibility of controlling the Fermi level via gating. The procedure, commonly referred to as the seeding-layer technique \cite{saito2015self,collins2015growth,2b4db284486745bb82d05e9bdd359f8e,}, involves a two-step growth where a small layer of amorphous material ---the seeds--- is initially deposited and subsequently annealed before continuing the high-temperature growth of the desired material. The Van der Waals epitaxy of MST on seeds of Sb\textsubscript{2}Te\textsubscript{3} using Pulsed Laser Deposition [PLD] is confirmed using atomically resolved STEM.
\section{Synthesis and Structure}
\subsection{Polycrystalline sample}
A polycrystalline ingot of MST was synthesized using a method adapted from Orujlu et al.\cite{orujlu2021phase}, where the binary compounds MnTe and Sb\textsubscript{2}Te\textsubscript{3} are used as precursors. The procedure involves an initial high-temperature homogenization phase, followed by quenching, and is concluded with a long annealing period ($\sim$2 months) at 600°C to reach the equilibrium state. These steps are necessary to obtain the MnSb\textsubscript{2}Te\textsubscript{4} phase since it is not directly accessible through the molten liquid due to a peritectic decomposition (MnSb\textsubscript{2}Te\textsubscript{4} $\to$ MnTe + L) at 647°C \cite{orujlu2021phase}. \textbf{Figure 1b} presents a micrograph of the ingot's surface taken with a scanning electron microscope [SEM] operated at 30kV using a Circular Backscattered [CBS] detector. As can be seen, the sample is composed of a compact network of fairly large crystallites of a couple hundreds of \textmu m with sharp angular features. The uniform contrast in the CBS image indicates similar composition thought the sample.\vspace{15pt}

The complete incorporation of the precursors is confirmed in \textbf{Figure 1d}, where the X-ray diffraction [XRD] spectrum of the sample's powder is compared with the signals expected for Sb\textsubscript{2}Te\textsubscript{3} \cite{anderson1974refinement,Downs2003,Grazulis2009} (COD ID: 9007590) and MnTe \cite{kruger1967lattice,Downs2003,Grazulis2009} (COD ID: 1539472). The spectrum exactly matches the one reported by Orujlu et al.\cite{orujlu2021phase} for MST and does not show the presence of Sb\textsubscript{2}Te\textsubscript{3}  nor MnTe. Moreover, the presence of additional phases like MnSb\textsubscript{4}Te\textsubscript{7} is also excluded through the small-angle scan. The analysis was repeated for multiple parts of the ingot to confirm its uniformity \textbf{(Note S1, Supporting Information)}. \vspace{15pt}

A lamella was prepared from the ingot by selecting a suitable oriented grain to display the vdW-like gaps in edge-on orientation. \textbf{Figure 1e} presents an inverse pole figure of the sample obtained with electron backscatter diffraction [EBSD], revealing that even after intense grinding, the exposed grains remained large (tens of microns). \textbf{Figure 1c} shows the cross-sectional image of the selected grain captured by high-angle annular dark-field [HAADF] STEM. The image shows a nearly perfect arrangement of SL separated by the vdW-like gap. Furthermore, due to the contrast mechanism of a HAADF detector, where atoms with a lower atomic number appear darker, the atomic arrangement of Mn within a SL can be clearly resolved as shown in the inset of \textbf{Figure 1e}. Detailed structural analysis \textbf{(Note S2, Supporting Information)} reveals that the contrast variation within a single Mn plane, which was interpreted by Y. Liu et al. \cite{liu2021site} as indicative of Mn-Sb site mixing, is absent from our specimen.\vspace{15pt}

Our sample presented few extraneous quintuple layers [QL] or even stacking faults, a common occurrence in layered tellurides \cite{chen2020ferromagnetism,levy2022compositional,levy2023high,forrester2024structural,momand2015interface,momand2016atomic}. Whenever observed, they would be concentrated near three-dimensional voids (see \textbf{Figure 2c}). This observation highlights the role that vacancy ordering plays during crystal formation. Indeed, in analogy with other GeSbTe-based structures, the vacancies likely became mobile during the high-temperature annealing, reconfiguring using the bi-layer defects until reaching the equilibrium MST structure; thus explaining the concentration of these defects solely around the voids \cite{momand2017dynamic}. 

\subsection{Thin film}
MnSb\textsubscript{2}Te\textsubscript{4} films were grown using PLD over thick ($\sim$300nm) amorphous SiO\textsubscript{x} aided by Sb\textsubscript{2}Te\textsubscript{3} seeds. The successful Van der Waals epitaxy is both confirmed by the preservation of the Reflection High-Energy Electron Diffraction [RHEED] pattern during growth \textbf{(Note S3, Supporting Information)} and by the cross-sectional HAADF-STEM images of \textbf{Figure 2a}. Furthermore, the HAADF intensity contrast (\textbf{Figure 2b}) confirms that the SLs in the structure are in fact MST with its central Mn planes excluding a possible metastable Sb\textsubscript{4}Te\textsubscript{3} phase \cite{yimam2023van}.\vspace{15pt}

The MST film had an approximate thickness of 23nm and was composed of small ($\sim$20nm) randomly in-plane oriented crystallites but with highly-oriented c-axis pointing out-of-plane, a standard signature of the seeding layer technique \cite{2b4db284486745bb82d05e9bdd359f8e,97dbe2aa85bc4d6c82158670cfb9b1e1,bc31de31519b4472a25d444fff03295c}. These reduced grains created many boundaries and defects, the most abundant of which were bi-layer defects. Frequently, these stacking faults created a complicated pattern in which the SL morphed into QL and even nonuple layers [NL] (see \textbf{Figure 2d}). Another common occurrence was the presence of twins and thus twin boundaries (\textbf{Figure 2e}); while twinning was observed in both SL and QL, it seemed to occur at a higher frequency between QLs. Moreover, since our samples were exposed to air, the upper surface was exclusively populated with QL, followed by a $\sim$5nm surface oxide. \vspace{15pt}

The need for heteroepitaxy was dictated by the failure to directly use MST as a seeding layer \textbf{(Note S3, Supporting Information)}. Indeed, the disappearance of RHEED patterns upon annealing strongly suggests that the thin amorphous MST film evaporates before crystallizing, thereby hindering the formation of seeds. In addition (see \textbf{Section 3}), the presence of oxygen at the surface stimulated Mn diffusion, further disrupting the crystal formation. Thus, Sb\textsubscript{2}Te\textsubscript{3} offers a proven \cite {yimam2023van,vermeulen2018strain} lattice-matched alternative for the growth of many tellurides over SiO\textsubscript{x}, bearing a strong resemblance to the procedure used by Tamargo et al. \cite{levy2022compositional,levy2023high,forrester2024structural} for fabricating Sb\textsubscript{2}Te\textsubscript{3}-MnSb\textsubscript{2}Te\textsubscript{4} multilayer. \vspace{15pt}

The final thickness of the film was limited by the appearance of ring-like features in the RHEED pattern \textbf{(Note S3, Supporting Information)}, indicative of nanocrystal formation whose increased roughness hindered further epitaxy. \textbf{Figure 2f} shows a cross-sectional HAADF-STEM image of such a nanocrystal. The structure matches the peritectic decomposition of MST into a liquid Sb\textsubscript{2}Te\textsubscript{3}-rich phase, and MnTe \cite{orujlu2021phase}, thus confirming that the nanocrystals are not grown on the film but rather transported by the plume after laser ablation. Indeed, the unfortunate presence of particulates on the film surface is a well-reported drawback of PLD \cite{barrington2000effect,yimam2021pulsed}; which, nevertheless, could be mitigated by different ablation protocols \cite{yimam2021pulsed,schenck1998particulate}.

\section{Elemental composition}
\subsection{Polycrystalline sample}
Having confirmed the structure of MST, the elemental distribution was studied with Energy-Dispersive X-ray Spectroscopy [EDS]. The composition of the ingot was confirmed to be uniform and of correct stoichiometry  [atomic percentage (error): 14.71\% (5.86\%) Mn, 28.03\% (2.98\%) Sb, 57.26\% (3.07\%) Te]. No additional phases were identified \textbf{(Note S4, Supporting Information)}. \vspace{15pt}

The atomic resolution elemental map of a single grain is shown in \textbf{Figure 3a}. The atomic planes and vdW-like gap are clearly resolved. \textbf{Figure 3b} presents their integrated profile. No intermixing is detected on the Te planes. In contrast, in the Mn and Sb planes, and even within the vdW gap, we observe a signal originating from both elements, thereby directly confirming Mn-Sb site mixing and Mn/Sb-interstitials by atomic resolution EDS-STEM. Unexpectedly, the Sb planes show a consistent asymmetry within a single SL: \textbf{Figure 3b} shows the Mn signal is higher at the "top" of the SL, while the Sb signal follows the opposite trend. This asymmetry occurs in all SL and is seen across much of the grain \textbf{(Note S5, Supporting Information)}. Furthermore, unlike observed in the samples by  Y. Liu et al. \cite{liu2021site}, our specimen has fully homogenized atomic planes, thus showing no compositional nor contrast variation perpendicular to the [0001] direction, i.e. in-plane \textbf{(Note S2, Supporting Information)}. \vspace{15pt}

While asymmetric structures are frequently encountered in crystals synthesized in out-of-equilibrium conditions, like those encountered during molecular beam epitaxy [MBE] \cite{momand2016atomic,levy2023high} or chemical vapor deposition [CVD]\cite{zou2024unusual}, a kinetic origin of our structure is highly unlikely given our sample's long anealing time ($\sim$2 months). Therefore, unlike previously believed \cite{liu2021site}, MST can actually develop an inversion-symmetry-breaking anti-site distribution at thermodynamic equilibrium. \vspace{15pt}

This surprising observation bears a strong resemblance to the recently proposed class of Janus materials \cite{chen2024screening,zou2024unusual,jiang2021topological} and thus could share both a similar origin and properties. For instance, the asymmetric structure of Sb\textsubscript{2}TeSe\textsubscript{2} [Janus-Sb\textsubscript{2}TeSe\textsubscript{2}: Te-Sb-Se-Sb-Se, mp-8612], an analogous tetradymite TI \cite{yeh2020femtosecond}, is predicted to be more stable than its symmetric counterpart [$\alpha$-Sb\textsubscript{2}TeSe\textsubscript{2}: Se-Sb-Te-Sb-Se, mp-571550] \cite{jain2013commentary,jain2011formation,chen2019new}. This asymmetry, however, has not been directly confirmed, in contrast with Janus-Bi\textsubscript{2}TeSe\textsubscript{2} \cite{kupers2020preferred, teramoto1961relations,Antwerpen,zou2024unusual}. Nonetheless, courtesy of the lower symmetry, Janus structures have been shown to exhibit phenomena not allowed in centrosymmetric structures, such as the Rashba effect\cite{zhang2021two}, polarization-dependent second-harmonic generation\cite{zou2024unusual}, and piezoelectricity\cite{qiu2021piezoelectricity}, among others. 
If similar properties will also be present in MST, which asymmetric structure we have directly demonstrated here, it could offer a platform for combining not just magnetism but also topology with effects like piezoelectricity and non-linear optics, offering a unique platform for topological devices and thus warrants further theoretical and experimental exploration.

\subsection{Thin film}
The elemental distribution of the MST thin-film is shown in \textbf{Figure 3c} with their respective integrated profiles in \textbf{Figure 3e}. In contrast with the ingot's single grain, these measurements had a much bigger background contribution due to the small crystallite size; nevertheless, the presence of Mn in the middle of the septuple layers could be directly confirmed with EDS, consistent with the HAADF results (\textbf{Figure 2b}), while the asymmetry in the Sb planes was not observed \textbf{(Note S5, Supporting Information)}. This feature is likely missing due to the out-of-equilibrium conditions involved during film growth. \vspace{15pt} 

Since the film was exposed to ambient conditions, a $\sim$5nm amorphous oxide layer developed at the upper surface. The film was exposed for several days; thus, the layer is presumably self-limiting. A detailed analysis of the different sections of the film (\textbf{Figure 3d}) reveals that the top section is Mn-rich and Te-deficient, thus explaining the overabundance of QL at the surface (\textbf{Figure 2e}). Indeed, Mn is the element with the strongest oxygen affinity in the compound \cite{moltved2019chemical} and thus out-diffuses to form the oxide, arguably making the SL unstable. In this regard, the Sb\textsubscript{2}Te\textsubscript{3} seeds had an additional effect by preventing the SiO\textsubscript{x} from reacting with MST, thereby enabling the film growth \textbf{(Note S3, Supporting Information)}.\vspace{15pt} 

Finally, \textbf{Figure 3d} presents a comparison of the stoichiometry of the film's core and the ingot, as determined by STEM EDS. There is a strong agreement between the measured percentages, consistent with the well-documented stoichiometric transfer property of PLD \cite{rijnders2006situ}. Although the data suggest the film may be Te-deficient, the exact stochiometry remains within the quantification confidence margins. Nevertheless, a more precise analytical method is recommended to further investigate potential Te loss. Indeed, as presented in other reports \cite{2b4db284486745bb82d05e9bdd359f8e,orgiani2017structural,noro1993thermoelectric}, chalcogenides' high vapor pressure promotes re-evaporation and material loss during ablation. This may inadvertently favors Mn-Sb intermixing, as Sb must also replace Te in the structure, presumably increasing the p-character of the film. The additional doping, however, might still be compensated by a potential gating enable by the Si-compatible seeding layer.


\section{Magnetic properties}
\subsection{Polycrystalline sample}
To complete the characterization, the magnetic properties of the polycrystalline powder were studied using DC-SQUID magnetometry. \textbf{Figure 4a} shows the isothermal magnetization curve at different temperatures. A small but clear hysteresis is observed at 5K which practically vanishes at 25K, in agreement with other reports on (ferri)ferromagnetic MST \cite{murakami2019realization,liu2021site,ge2021direct}. \vspace{15pt}

The magnetic moment of the sample under -5mT field cooling [FC] is plotted in \textbf{Figure 4d}, where a maximum is reached around 20K. The feature is highly sensitive to the cooling conditions and disappears at a moderate field strength \textbf{(Note S6, Supporting Information)}. While the presence of such a maximum is commonly linked to antiferromagnetic order \cite{blundell2001magnetism}, it has also been reported in (ferri)ferromagnetic samples of MnSb\textsubscript{1.8}Bi\textsubscript{0.2}Te\textsubscript{4} \cite{chen2020ferromagnetism} and in the in-plane (perpendicular to the crystallographic c-axis) response of MnSb\textsubscript{2}Te\textsubscript{4} single crystals\cite{liu2021site,xi2022relationship}. Given that in our randomly oriented powder sample the in-plane directions are overrepresented due to the plate-like nature of our specimen (see \textbf{Figure 1b}), we suspect that the maximum is in fact caused by the magnetocrystalline anisotropy rather than an antiferromagnetic order. \vspace{15pt}

To further elucidate the magnetic properties, the ratio between the applied field and the magnetic moment is depicted in \textbf{Figure 4c}. The quotient is inversely proportional to the susceptibility and thus is expected to follow the Curie-Weiss law above the transition temperature \cite{blundell2001magnetism}. The linear fit intersects the abscissa at a positive temperature near 25K, consistent with a (ferri)ferromagnetic sample. Nevertheless, when the ratio is measured at higher temperatures (see \textbf{Figure 4e}, where the field strength was increased to maintain a high signal-to-noise ratio), the linear trend changes to a smaller slope, thereby shifting the intersection to lower temperatures ($\sim$4K).\vspace{15pt}

The paramagnetic anomaly of MST has also been observed by T. Murakami et al.\cite{murakami2019realization}, and J.-Q. Yan et al. \cite{yan2019evolution}, irrespective of whether the sample exhibits ferri- or antiferromagnetism. The deviation occurs between 50-80K, and was suggested by J.-Q. Yan et al. \cite{yan2019evolution} to be caused by short-range magnetic correlations. Indeed, the higher slope at lower temperature (\textbf{Figure 4e}) implies a smaller Curie constant, which signals a "freezing" of effective spins \cite{blundell2001magnetism}. Noting that all reported MST samples, including ours, have an important amount of Mn-Sb site mixing, we suspect that a mechanism similar to the spin-glass transition of the hematite-ilmenite solid solution may be at play \cite{brown1993hematite}. In this system, the presence of Fe$^{3+}$ anti-sites in the Ti$^{4+}$ sublattice is believed to produce small regions of antiferromagnetic coupling between Fe$^{2+}$, thereby lowering the effective number of spins. By analogy, the presence of Mn anti-sites in the Sb planes, together with Mn interstitials within the vdW-like gap (directly observed in \textbf{Figure 3a}), could create equivalent antiferromagnetic pockets via a Mn-Te-Mn superexchange \cite{PhysRevB.88.235131}. Nevertheless, a detailed microscopic description still needs further development. \vspace{15pt}

The remanent magnetization [RM] after FC is plotted in \textbf{Figure 4d}. The RM curves exhibit several unusual features. For instance, when the inverse magnetization is plotted (see \textbf{Figure 4f}), the high-temperature anomaly reappears. Moreover, two distinct inflection points are evident in \textbf{Figure 4d}. Similar low-temperature traits,  significantly deviating from the behavior of an ideal ferromagnet \cite{blundell2001magnetism}, have also been reported in the literature \cite{wimmer2021mn,folkers2022occupancy}. To our knowledge, these features remain unexplained but may be linked with the small presence of Mn-doped Sb\textsubscript{2}Te\textsubscript{3} QL, as proposed by Isaeva et al. \cite{folkers2022occupancy} and Tamargo et al. \cite{levy2023high, forrester2024structural}, the latter in the context of Sb\textsubscript{2}Te\textsubscript{3}-MnSb\textsubscript{2}Te\textsubscript{4} multilayer. Such interpretation aligns well with our observations of QL concentrated near defects, even in seemingly perfect crystals (see \textbf{Figure 2c}). \vspace{15pt}

Overall, the observed hysteresis loop and RM provides strong evidence that our sample exhibits ferrimagnetism, consistent with the findings of T. Murakami et al. \cite{murakami2019realization} for polycrystalline MST. Our direct atomic-scale observations reveals extensive Mn site mixing, supporting a structure of ferromagnetic Mn planes antiferromagnetically coupled to Mn anti-sites in Sb planes, as described by Y. Liu et al. \cite{liu2021site}. An often overlooked factor in this context is the presence of Mn interstitials, which we have also identified in our long-annealed sample (see \textbf{Figure 3b}) and appear to be thermodynamically favored due to increased configurational entropy. Indeed, their presence introduces an additional coupling mechanism that may further enhance ferromagnetic interactions between intra-septuple Mn planes, similar to the behavior observed in other Mn-rich compounds (e.g. Mn\textsubscript{2}Sb\textsubscript{2}Te\textsubscript{5}) as reported by Isaeva et al.\cite{kochetkova2025mn}.

\subsection{Thin film}
The DC-SQUID magnetometry results for the MST thin film are plotted in \textbf{Figure 4b}. Significant challenges were encountered during the measurements of the thin film sample due to the competing diamagnetic interaction of the Si/SiO\textsubscript{x} substrate which severely diminished the total signal. Nevertheless, a clear distinction could still be observed between the out-of-plane (perpendicular to the film) and in-plane response. In neither case, a hysteresis could be observed due to high noise levels. \vspace{15pt}

Courtesy of the seeding layer technique, the out-of-plane direction coincides with the crystallographic c-axis. Contrary to other reports\cite{wimmer2021mn,liu2021site}, our in-plane response was considerably stronger than the out-of-plane. These observations point to a significant contribution from surface and interface anisotropy \cite{fernando2008magnetic}, likely due to the reduced film thickness and the numerous grain boundaries. Overall, the measurement technique is inadequate to disentangle these contributions; thus, other approaches more suited to the study of magnetization in ultra-thin samples, such as Anomalous Hall measurements\cite{yan2019evolution}, should be employed and will be the focus of future research.


\section{Conclusion}
The extraordinary sensitivity of MST to atomic-scale defects establishes atomic structure studies as fundamental for unlocking the mechanisms governing the interplay of topology and magnetism. Such insights not only propel our understanding of MST, but also redefine design principles for next-generation quantum materials. In our work, we have demonstrated that the anti-site distribution of MST can develop, in equilibrium conditions, an anisotropy that clearly distinguishes between the two Sb planes within a SL, while still exhibiting all of its well-reported ferrimagnetic properties \cite{murakami2019realization}. The site mixing is not a universal feature of the compound and depends strongly on synthesis conditions; our research suggests that it develops during long high temperature annealing and is suppressed in out-of equilibrium conditions. The observation of such a structure should warrant further studies and a reevaluation of previous assumptions. For instance, its omission in the expected XRD analyses could misestimate the actual Mn-Sb site mixing of samples. Moreover, it could be an important piece of information for theories aiming to explain MST's magnetic anomalies. In addition, the inversion-symmetry-breaking suggests MST as a prospective member of the emerging class of Janus materials, potentially combining magnetism, topology, and piezoelectricity, among other effects, in a single material. \vspace{15pt}

Lastly, we presented the successful Van der Waals epitaxy of MST over Sb\textsubscript{2}Te\textsubscript{3}, allowing for its thin film growth over SiO\textsubscript{x}. Although the seeding-layer approach, has a long history in the growth of chalcogenides-based structures for applications in Phase Change Memory [PCM] and thermoelectrics \cite {2b4db284486745bb82d05e9bdd359f8e,yimam2023van,vermeulen2018strain,97dbe2aa85bc4d6c82158670cfb9b1e1,bc31de31519b4472a25d444fff03295c,nie2025role}, it has been considerably underexplored in the context of quantum materials, even though its successful use to grow film of materials like the quintessential Bi\textsubscript{2}Se\textsubscript{3}, with persistent topological properties, has already been demonstrated \cite{bansal2014robust}. The enhanced Si compatibility and the possibility of controlling the effective dopant level via gating, rather than fine-tuning deposition parameters, represent a significant stepping-stone in developing devices based on topology and will undoubtedly play an important role in future research studying the role of carriers in the magnetic and topological properties of MST. \vspace{15pt} 


\section{Experimental Section}
\threesubsection{Solid Synthesis}\\
Sb\textsubscript{2}Te\textsubscript{3} was pre-synthesized by mixing 6.490gr of high-purity Sb (99.999\% Alfa Aeser) and 10.203gr of Te (99.999\% Sigma Aldrich) in an evacuated quartz ampule which was heated up to 700°C (at 5°C/min) in a Nabertherm Box Furnace and left to equilibrate at that temperature for 7hrs followed by a slow cool down to room temperature at a maximum rate of 5°C/min. A small amount of powder was analyzed with XRD (as described below) to confirm the formation of the correct crystal structure (cross-checked with COD ID 9007590 \cite{anderson1974refinement}).

The MnSb\textsubscript{2}Te\textsubscript{4} polycrystalline ingot was synthesized in a two-step procedure using MnTe (99.99\% Stanford Advanced Materials) and the pre-synthesized Sb\textsubscript{2}Te\textsubscript{3} as precursors. The method was adapted from the synthesis presented by Orujlu et al. \cite{orujlu2021phase}. Firstly, 3.396gr of MnTe and 11.647gr of Sb\textsubscript{2}Te\textsubscript{3} were mixed in a carbon-coated quartz ampule (Sandfire Scientific Ltd.) to avoid any adverse reaction of the Mn with the ampule \cite{orujlu2021phase,canaguier2020kinetics}. The evacuated ampule was heated to 900°C at 1.7°C/min and left for 24hr to homogenize followed by a quenching in icy-water. Secondly, the ampule was annealed at 600°C for 62 days to reach its equilibrium state. Finally, the ampule was cooled down to room temperature at a maximum rate of 2°C/min. 

\threesubsection{Characterization using XRD}\\
To obtain the X-ray diffraction [XRD] patterns, both a Bruker D8 Advance and a Bruker D8 Endeavor diffractometers using CuK$_\alpha$ radiation at room temperature in the Bragg-Brentano geometry were employed. The powder samples were laid on a Si low background sample holder, and the spectra were recorded between 2$\theta$= 15°÷75° (5°÷20°) for 1 sec/measurement (3 sec/meas) at a resolution of 3506 (752) measurements. Many powder samples from different parts of the MnSb\textsubscript{2}Te\textsubscript{4} ingot were measured to confirm uniformity. In addition, the spectrum of the empty sample holder was also recorded to account for environmental peaks.

\threesubsection{Characterization using SEM, EDS, and EBSD}\\
To expose a flat surface for imaging and elemental mapping, the MnSb\textsubscript{2}Te\textsubscript{4} ingot was ground using P500, P1000, P2400, and P4000 sandpaper (in that order) with ethanol. In addition, a small piece of the ingot was removed with a blade to be used for lamella preparation. The piece was extensively polished using Ar ion polishing (Gatan PIPS II Model 695); a series of 5kV polishing for 3 hours, 3kV for 2 hours, and 1kV for 1 hour were repeated 4 times to obtain indexable Kikuchi patterns in the Electron Backscattered diffraction [EBSD] signal.
The micrographs were taken with a FEI NovaNanoSEM 650 Scanning Electron Microscope [SEM] equipped with a Concentric Backscattered [CBS] and EDAX detector. The elemental composition and uniformity of the target were measured via Energy-Dispersive X-ray Spectroscopy [EDS] by collecting the spectrum from a 1800 $\times$1400 \textmu m$^2$ at a resolution of 512$\times$400 pixels for about 6 hours with a 10 kV beam with a dwell time of 200 \textmu s and 3.84 \textmu s amplification time. The spectral deconvolution was performed with the EDAX TEAM software (version 4.6.2004.0297), using the eZAF mode for quantification.
Finally, the Kikuchi patterns were collected from a 44$\times$69 \textmu m$^2$ area with a step size of 0.4 \textmu m using a 20 kV beam. The indexing and inverse pole figure were obtained using the OIM Analysis software (version 8.1.0); the latter was performed after a standard grain dilation clean-up procedure.

\threesubsection{Growth using PLD}\\
The thin films were grown on a 5$\times$5 mm$^2$ boron-doped Si wafer piece with 300nm of dry thermal oxide (University wavers ID: 3594). Before every deposition, the substrates were cleaned by sonicating in isopropanol [IPA] for 15 min at 30°C, followed by an IPA flush and drying with a N$_2$ gun. The samples were glued to the heater in the PLD with a small drop of Ag paint.
All depositions were done in a (TSST) Pulsed Laser Deposition [PLD] system using a KrF excimer laser (wavelength of 248nm) with a high-pressure 30 kV Reflection High-Energy Electron Diffraction [RHEED] system for in-situ monitoring. For the Sb\textsubscript{2}Te\textsubscript{3}-MnSb\textsubscript{2}Te\textsubscript{4} films, a laser fluence of 1 J cm$^{-2}$ was established before growth started; in addition, a repetition rate of 1 Hz, a target-to-substrate distance of 5.5mm and a processing gas (Ar at 1 sccm) pressure of 0.12mbar were used. The films were grown following the seeding layer procedure \cite{2b4db284486745bb82d05e9bdd359f8e}; firstly, 150 pulses of a vacuum-sintered powder target of Sb\textsubscript{2}Te\textsubscript{3} (99.99\% KTech supplies) were deposited at room temperature to create a thin (3-4nm) layer. Then, the seeds were heated up to 210°C at 5°C/min. After 10 minutes, the growth continued with 3000 pulses of MnSb\textsubscript{2}Te\textsubscript{4} using the synthesized ingot as a target. The sample was then cooled down to room temperature in the same Ar atmosphere. The samples were stored in ambient conditions for 22 days before preparation of a lamella.

\threesubsection{Lamella preparation using FIB}\\
Electron-transparent samples were prepared in a FEI Helios G5 CX dual-beam SEM-FIB system. A protective C layer, then a Pt layer, was deposited by the electron beam in all cases. This was followed by a thick, ion beam-induced Pt protective layer. A cross-sectional chunk, about 15$\times$2 $\times$5 \textmu m$^3$ was extracted and transferred with a W needle to a Cu half-grid. The chunks were then thinned to a final thickness of 80–100 nm using a Ga-beam at 30kV. For all samples, a rigid frame was left to minimize bending and release stress of the electron-transparent window. Finally, several low kV polishing steps (5–2 kV) were applied to remove damage layer and clean the transparent window.

\threesubsection{STEM imaging and STEM-EDS analysis}\\
The lamellas were mounted on a dedicated double-tilt holder optimized for X-ray collection. Atomic-resolution images were captured with a double-corrected and monochromated (Thermo Fisher Scientific) Themis Z Scanning Transmission Electron Microscope [STEM] operating at 300 kV and equipped with a High-Angle Annular Dark-Field [HAADF] STEM detector, and a 4-segmented (integrated) Differential Phase Contrast [iDPC] annular bright field detector. The beam convergence angle used was ~24.0, and ~35mrad. A probe current of 20pA was used for imaging.
The Energy Dispersive X-ray Spectroscopy [EDS] were measured with a Bruker Dual X EDS system consisting of two large area detectors capturing 1.76 steradian with a probe current of 100pA.
The analysis of all STEM images and STEM EDS maps were carried out using the Velox software (version 3.16.1.500). To estimate the stoichiometry of the ingot using STEM EDS, small samples of powder were dispersed in a Cu grid covered with a thin amorphous carbon membrane. Once an electron-transparent particle was found, the elemental map was collected. The process was repeated 5 times to obtain confidence intervals for the stoichiometry. For quantification, all the deconvoluted spectra were analyzed using the Brown-Powell ionization cross-section model after a multi-parabolic background correction.

\threesubsection{Characterization using SQUID}\\
The magnetic properties of the samples were measured using a Quantum Design Magnetic Property Measurement System [MPMS] XL equipped with a DC-superconductor quantum interference device [SQUID] magnetometer. The powder sample was inserted on a plastic capsule with cotton as support. All samples were suspended from the inserting rod using a plastic straw and Kapton tape. The samples were moved over 6.30cm, collecting 64 points per scan. The average of three scans was used for fitting the magnetic moment to the recorded signal. For magnetization vs field curves, the samples were first cooled down to the desired temperature, followed by a saturation of the magnetic moment at a field of +5T before starting the measurements. 

\medskip
\textbf{Supporting Information} \par 
Supporting Information is available from the Wiley Online Library or from the author.

\medskip
\textbf{Acknowledgements} \par 
We would like to thank Jacob Baas, Elnur Orujlu, Maxim Mostovoy, and Antonija Grubisic-Cabo for useful discussions; and Javier Chavez Ponce de Leon for his helpful assistance in the preparation of the thumbnail. This research was supported by the Netherlands Organizationf for Scientific Research (NWO) under Grant No. OTP 19527.

\medskip
\textbf{Conflict of Interest} \par 
The author declare no conflict of interest.

\medskip
\textbf{Data Availability Statement} \par 
The data that supports the findings of this study are available from the corresponding author upon reasonable request.

\medskip

%
\printbibliography

\begin{figure}
  \includegraphics[width=\linewidth]{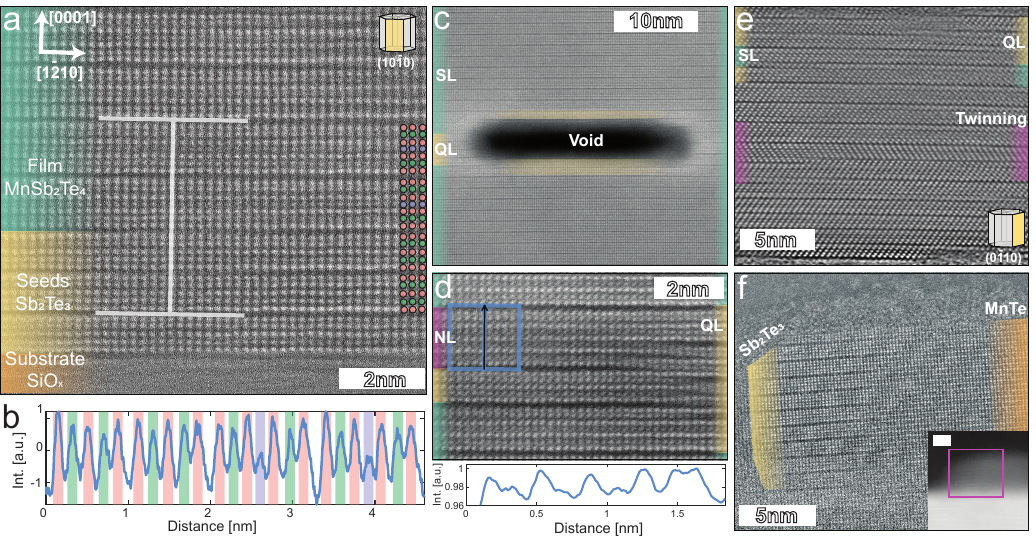}
  \caption{Structure and defects of MnSb\textsubscript{2}Te\textsubscript{4}. a) HAADF STEM images of the thin film. SL of MnSb\textsubscript{2}Te\textsubscript{4} are epitaxially grown over three QLs of Sb\textsubscript{2}Te\textsubscript{3} seeds over amorphous SiO\textsubscript{x}. b) Intensity profile of the area marked in (a), the contrast in the HAADF image is generated by difference in atomic number enabling the identification of Mn in the structure. The Sb peaks show slight contrast variation from the Te suggesting Mn-Sb site mixing. c) HAADF STEM image of a single MnSb\textsubscript{2}Te\textsubscript{4} grain. The only defects found in the grain were 3D voids where occasional QL (in yellow) could be found. d) HAADF STEM image with a Radial Wiener filter of the MnSb\textsubscript{2}Te\textsubscript{4} thin film. Abundant bilayer defects are present which mediate the position of QL and SL in the film. Occasionally they even combine to form nonuple layers [NL]. Its corresponding intensity profile is plotted below. e) iDPC STEM image of the MST film showing a twin boundary. Due to oxidation, only QL can be bound at the upper surface of the film. f) HAADF STEM image of a particulate in the surface of the film transported during pulsed laser deposition. The particulate shows the peritectic decomposition of MnSb\textsubscript{2}Te\textsubscript{4} into MnTe and Sb\textsubscript{2}Te\textsubscript{3}.}
  \label{fig:Structure}
\end{figure}

\begin{figure}
  \includegraphics[width=\linewidth]{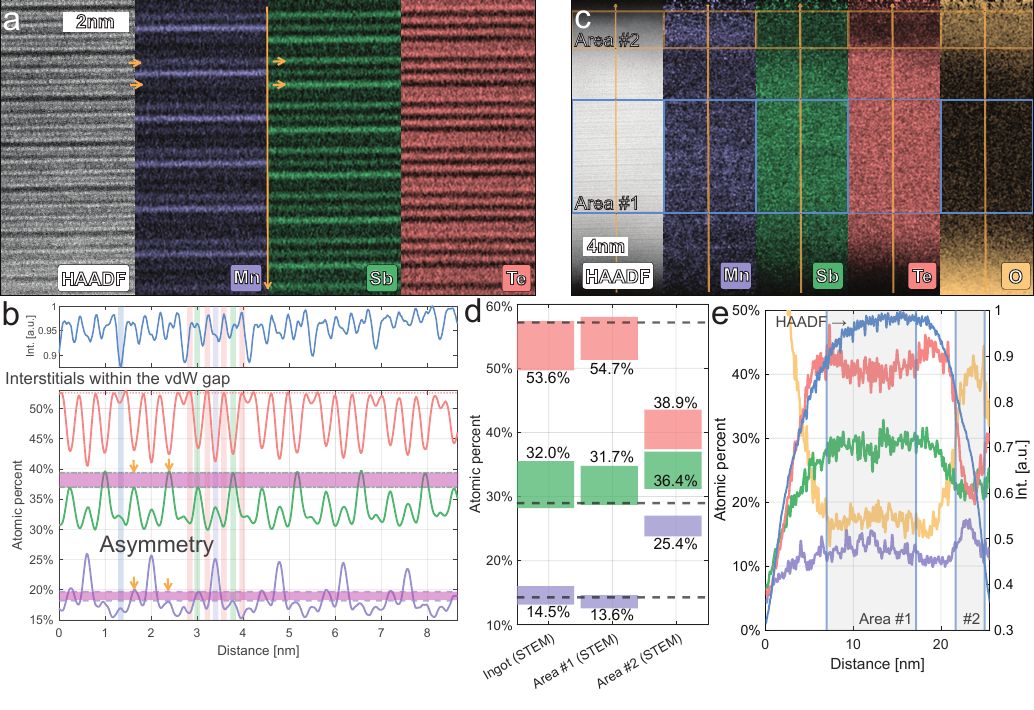}
  \caption{Stichometry and atomic distribution in MnSb\textsubscript{2}Te\textsubscript{4}. a) Atomically resolved elemental map of a single MnSb\textsubscript{2}Te\textsubscript{4} grain measured with EDS in STEM mode (integration over 5581 frames). The Mn-Sb anti-sites are asymmetrically distributed. b) Intensity profiles corresponding to Fig. (a) with integration width of 16 nm. A significant Mn signal can be found in the Sb planes and even within the Van der Waals gap as interstitials. The orange arrows indicate the asymmetry in the Sb planes with one plane showing more Mn signal and correspondingly less Sb. The asymmetry occurs within a septuple layer and it is consistent in all septuple layers and multiple parts of the grain. No asymmetry in the Te planes is detected. c) Elemental distribution within the thin film. In the presence of oxygen, Mn diffuses from the center of the film (area \#1) to the surface (area \#2) creating a Mn-rich amorphous oxide. d) Stoichiometry of the ingot (as determined by EDS in STEM mode), the bulk of the film, and the amorphous oxide (areas \#1 and \#2 in Fig. (c) respectively). e) Line profile of Fig. (c). The HAADF signal is plotted in blue and corresponds to the right-side axis. The increase of Mn and O near the surfaces is clearly visible.}
  \label{fig:Stochiometry}
\end{figure}

\begin{figure}
  \includegraphics[width=\linewidth]{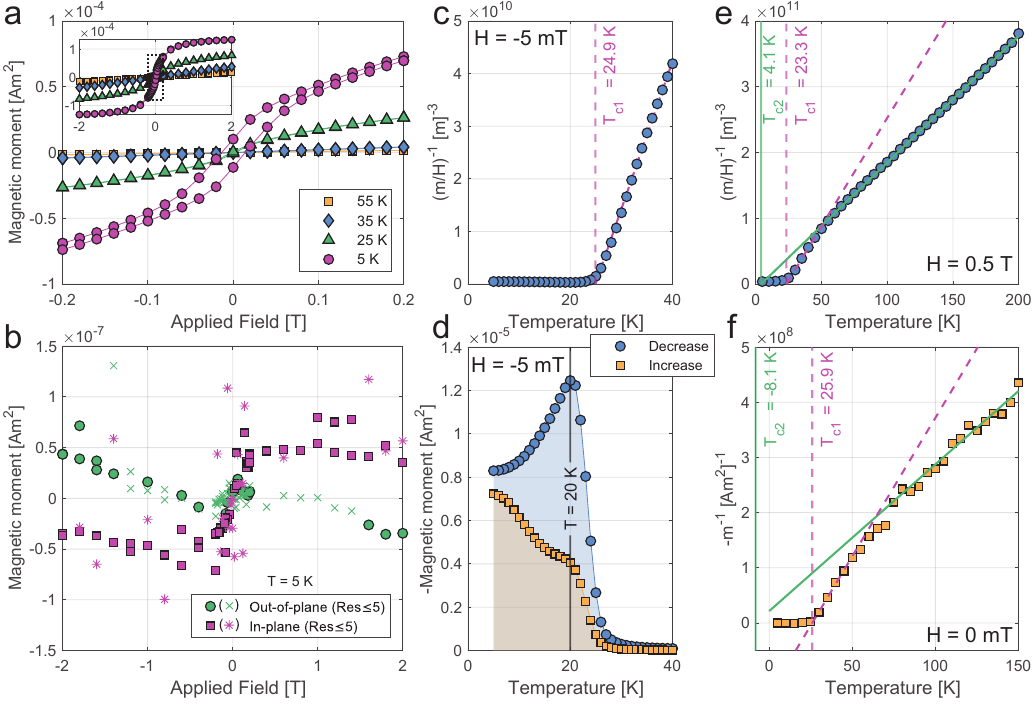}
  \caption{Magnetic properties of MnSb\textsubscript{2}Te\textsubscript{4}. a) Magnetic moment vs applied field (m vs H) loops at different temperatures of MnSb\textsubscript{2}Te\textsubscript{4} powder. A clear hysteresis is present at 5 K indicating ferri/ferromagnetic order. b) m vs H loops at 5 K of the MnSb\textsubscript{2}Te\textsubscript{4} thin film for different direction. The signal is poor due to the small film thickness and the diamagnetic response of silicon. The data point with bad residual are plotted with crosses and asterisks. Anisotropy is clearly observed, but no hysteresis can be discerned. No background signal has been removed. c)  H/m plot during a -5mT FC of the powder sample. The plot is proportional to the inverse susceptibility and exhibits a Curie-Weiss behavior (straight line fit) with critical temperature near 25 K. d) FC and RM at -5mT of the powder sample. During FC, the magnetic moment develops a maximum near 20 K reminiscent of antiferromagnetic order. The RM is always decreasing, but has two inflection point. e) High-temperature H/m plot during a 500mT FC of the polycrystalline sample. The Curie-Weiss behavior changes slope around 50-70 K. f) High-temperature RM after the 500mT FC.}
  \label{fig:MagneticProperties}
\end{figure}

\begin{figure}
  \includegraphics[width=\linewidth]{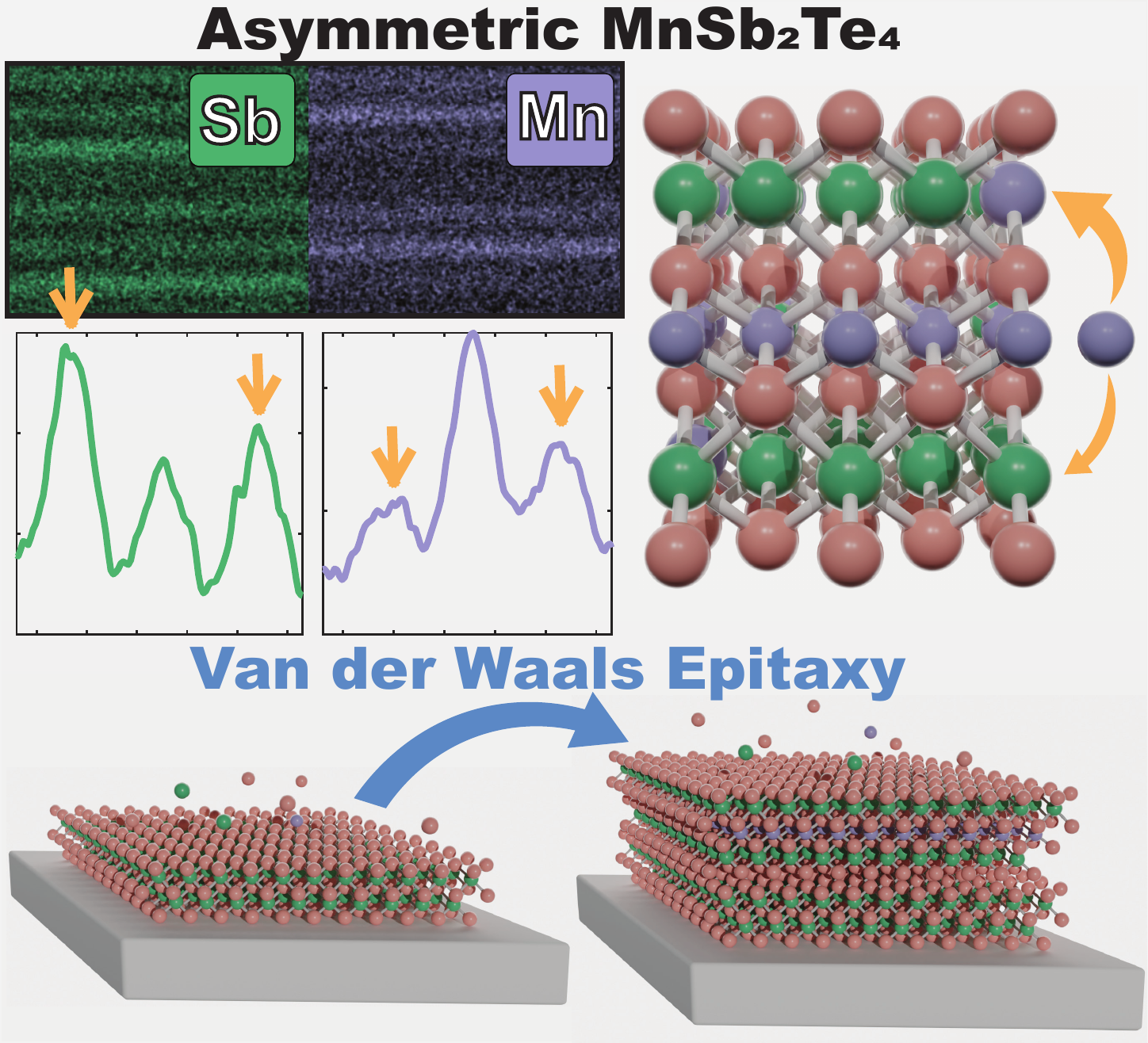}
  \caption{Table of content/Thumbnail figure}
\end{figure}

\foreach \x in {1,...,16}{
\begin{figure}
\vspace{-60pt}
  \includegraphics[page=\x]{SupplementaryMaterialVersion2.pdf}
\end{figure}
}

\end{document}